\documentclass[preprint,tightenlines,showpacs,preprintnumbers,amsmath,amssymb,prb]{revtex4}

\usepackage{graphicx}
\usepackage{bm}

\begin{document}

\title{ Electronic and Transport Properties of Radially Deformed Double-walled Carbon Nanotube Intramolecular Junction}
\author {Xiaoping Yang}

\affiliation{Group of Computational Condensed Matter Physics,
National Laboratory of Solid State Microstructures and Department
of Physics, Nanjing University, Nanjing 210093, P. R. China}
\affiliation{Department of Physics, Huainan Normal University,
Huainan, Anhui 232001, P. R. China}
\author {Jinming Dong}
\affiliation{Group of Computational Condensed Matter Physics,
National Laboratory of Solid State Microstructures and Department
of Physics, Nanjing University, Nanjing 210093, P. R. China}

\begin{abstract}

The electronic and transport property of a radially deformed
double-walled carbon nanotube (DWNT) intramolecular junction (IMJ)
has been studied by the tight-binding (TB) model combined with the
first-principle calculations. The geometrical structures of the
DWNT IMJ have been first optimized in energy by the universal
force field (UFF) method. It is found that when heavily squashed,
the DWNT will become an insulator-coated metallic wire, and the
conductance near the Fermi level has been significantly changed by
the radial squash. Specially, several resonance conductance peaks
appear at some energies in the conduction band of the squashed
DWNT IMJ. Finally, we have also investigated the conductance
variation due to change of the length of the central semiconductor
in the squashed DWNT IMJ. Furthermore, a promising pure carbon
nanoscale electronic device is proposed based on the DWNT IMJ.

\end{abstract}


\maketitle
\section{Introduction }

In the past decade carbon nanotubes (CNTs) [1-3], both
single-walled (SWNT) and multi-walled (MWNT) had been extensively
investigated due to their special electrical and mechanical
properties, as well as their potential applications in future
nanostructured materials, such as nanoscale quantum wires, single
electron and field-effect transistors and sensors.

The SWNT is composed of a rolled-up 2D-graphite sheet, and
discovered first by Iijima's group in 1991. The carbon atoms on a
SWNT are arranged on a helical line around its axis. The
geometrical structure of a SWNT can be defined by a pair of
intergers ($n, m)$, which determines its radius and chirality, and
so entirely its electronic structure, and optical property. It is
known that the SWNTs with $n-m$ being a multiple of 3 are
metallic, and all others are semiconducting [4]. The DWNT is the
simplest type of MWNT, which has been made experimentally by many
different methods. The nucleation of its inner tube should occur
after the growth of the outer tube according to the 'yarmulke
mechanism' [5], which means that the inner tube diameter is
determined by that of the outer tube. A DWNT can be composed of a
pair of inner and outer constituent layers with any chiralities,
leading to different kinds of DWNT, such as metal--metal,
metal--semiconducting, or semiconducting--semiconducting
nanotubes. Early studies on DWNT focused mainly on their
electronic structure, and stability, etc. [6--9]. It was shown
that the band structure of a DWNT depends on the combined
configurations of the inner and outer tubes [7], but their
stability depends only on their interlayer spacing [6].

It is well known that a mechanical deformation of a SWNT affects
heavity its electronic properties [10--14]. For example, Lu et al.
[15,16] indicated that a metal-to-semiconductor transition (MST)
can be achieved by a radial deformation of the armchair SWNT.
Furthermore, they studied the transport properties of effectual
metal--semiconducting--metal (MSM) heterojunction in the metallic
armchair tube, and the effect of deformed radially finite length.
In addition, a natural MST is found theoretically in the DWNT
(7,0)@(16,0) and (7,0)@(17,0) due to the differences of the
downward shifts of the $\pi$ and $\pi ^ * $ electron states
between the inner and outer nanotubes [17], which reminds us of the
possible existence of the DWNT junction. Recently, a new
metal--semiconducting MWNT IMJ has been made [18], which shows
reproducible rectifying diode behavior. So, it is interesting to
study the electronic structure and transport property of the DWNT
IMJ, produced by squashing radially a DWNT, which is just the main
goal of this paper.

The paper is organized as follows: In Sec. II, we introduce the
model Hamiltonian, and the employed method. Then, the calculated
results and discussions are given in Sec. III. The conclusions are
shown in Sec. IV.

\section{Models and Method}

A finite DWNT (10,0)@(19,0) with a central segment of inner tube
uncovered is shown schematically in Figs. 1(a) and 1(b), viewed
from $x$-axis and $z$-axis direction, respectively. Two identical
lithium tips (with definite width of $d_{x}$ and length of
$d_{z}$) are applied from $\pm y$ directions on the system with a
separation distance $d_{y}$. With the two tips moving towards the
DWNT, the cross sections of the inner and outer tube are deformed
from circle to ellipse, and finally to stadium-shaped shown in
Figs. 1(c) and 1(d). Then the geometrical stucture of the DWNT for
various tip distances $d_{y}$ are optimized by the universal force
field (UFF) method [19,20]. When the deformation becomes bigger
enough, the inner (10,0) tubes in the left DWNT and the right one
will be able to exhibit a transition from semiconducting to metal.
In this case, the original system will become a MSM DWNT IMJ,
which has no any topological defects in the M-S and S-M junction
region.

After the geometrical stucture of the DWNT IMJ is optimized, we will
employ a TB Hamiltonian including four orbitals per atom to study
its electronic and transport properties, in which the $\sigma-\pi$
electron hybridization can be included [21,22,23]. The model
Hamiltonian can be written as follows:
\begin{center}
\begin{equation}
H=\epsilon_{s}^{0}c_{is}^{\dag}c_{is}+
   \sum\limits_{jp}\epsilon_{p}^{0}c_{jp}^{\dag}c_{jp}+
\sum\limits_{ijmn}(t_{ij}^{mn}c_{im}^{\dag}c_{jn}+H.c.),
\end{equation}
\end{center}
where $\epsilon_{s}^{0}$ and $\epsilon_{p}^{0}$ are the on-site
energies of the 2$s$ and 2$p$ orbitals, respectively.
$c_{is}(c_{is}^{\dag})$ and $c_{jp}(c_{jp}^{\dag})$ denote the
annihilation (creation) operators of an electron on Carbon 2$s$
orbitals at site $i$ and Carbon 2$p$ orbitals at site $j$,
respectively. $m$ and $n$ are the orbital indices. $t_{ij}^{mn}$
are the hopping integrals between the $m$ orbital on the site $i$
and the $n$ orbital on the first or second-nearest neighbor site
$j$, which are expressed in terms of Slater-Koster parameters
$V_{ss\sigma}$, $V_{sp\sigma}$, $V_{pp\sigma}$ and $V_{pp\pi}$
[24]. Since the inter-wall interaction strength is about eighth of
the intralayer hopping integrals, which has minor effect on the
band structure of a DWNT [25], so we do not take it into account
in this paper.

The parameter values in Eq. (1) are taken to be close to those
used for graphite in Ref. [26], which was also successfully used
to study the physical properties of the SWNT with small-radius
[23], the SWNT with polygonized cross sections [21]. Among the
four orbitals per atom, its $s$ level is located at
$\epsilon_{s}^{0}=-7.3$ eV below the triply-degenerated $p$ level
taken as the zero of energy ($\epsilon_{p}^{0}=0.0$ eV). The
Slater-Koster hopping parameters for the nearest-neighbor pairs
are taken as $V_{ss\sigma}=-4.30$ eV, $V_{sp\sigma}=4.98$ eV,
$V_{pp\sigma}=6.38$ eV and $V_{pp\pi}=-2.66$ eV. The
second-neighbor interactions are taken into account by using
$V_{ss\sigma}=-0.18Y$ eV, $V_{sp\sigma}=0.21Y$ eV,
$V_{pp\sigma}=0.27Y$ eV and $V_{pp\pi}=-0.11Y$ eV, where
$Y=(3.335/r_{ij})^2$ is a scaling factor, depending on the
interatomic distance $r_{ij}$ (in unit of \AA).

First, we study the band structures of the radially deformed DWNT
IMJ, and then, its transport properties . In the conductance
calculation, the whole system is considered to be composed of two
leads (left and right ones) plus a central DWNT IMJ (L-C-R), where
the two leads are taken as the same type of the deformed metal
inner tube (10,0) as those in the left and right deformed DWNT of
the system. This problem can be most conveniently treated by the
Green's function matching method [27], in which the conductance is
expressed by the Landauer formula:
\begin{center}
\begin{equation}
G=(2e^{2}/h)Tr[\Gamma_{L} G^{r}_{C}\Gamma_{R} G^{a}_{C}],
\end{equation}
\end{center}
where,
\begin{center}
\begin{equation}
G^r_{C}=(\epsilon+i\gamma-H_{C}-h^{\dag}_{LC}g_{L}h_{LC}-h^{\dag}_{RC}g_{R}h_{RC})^{-1},
\end{equation}
\end{center}
\begin{center}
\begin{equation}
G^a_{C}=(\epsilon-i\gamma-H_{C}-h^{\dag}_{LC}g_{L}h_{LC}-h^{\dag}_{RC}g_{R}h_{RC})^{-1},
\end{equation}
\end{center}
and
\begin{center}
\begin{equation}
\Gamma_{L,R}=ih^{\dag}_{L,RC}(g^{r}_{L,R}-g^{a}_{L,R})h_{L,RC}.
\end{equation}
\end{center}
Here, $H_C$ is Hamiltonian of the central DWNT IMJ, and $h_{L,C}$
($h_{R,C}$) is the coupling matrix between the central DWNT IMJ
and the left (right) lead. $g_{L}$ and $g_{R}$ are the Green's
functions of the semi-infinite left and right leads, which are
calculated by using an iterative procedure [28]. Also, we can
calculate the local density of states (LDOS) by using
LDOS$(j,\epsilon)=-(1/\pi)Im(G_{C}^{r}(j,j,\epsilon))$, where $j$
is the ordinal number of carbon atoms in the DWNT IMJ.

\section {Results and Discussions}

Based on the computational scheme, we optimized the geometrical
structure of a finite DWNT (10,0)@(19,0) IMJ under the different
tip distance $d_{y}$. The obtained results show that small radial
deformations generally cause elliptical cross sections for the
DWNT. A further squashed deformation will lead the tubes to be
flattened, producing a stadium-shaped cross section ($d_{y}=$
13.67 \AA, Fig. 1(c) and 1(d)), and meanwhile causing a
semiconductor-to-metal transition (SMT) for the inner (10,0) tube
of the DWNT. So, by using this model and the optimization
procedure, we can get any of the MSM or
semiconducting--metal--semiconducting (SMS) DWNT IMJ.

First, we study the electronic structure of the DWNT (10,0)@(19,0)
IMJ. It is found that the outer tube (19,0) always remain in the
semiconducting state when squashed (see Fig. 2(g)). But the band
gap of the inner tube (10,0) decreases significantly from 0.69 eV
to zero. Thus the squashed DWNT becomes an insulator-coated
metallic wire. The obtained band structures for the perfect (10,0)
tube and a series of deformed inner (10,0) tubes are shown in
Figs. 2(a)-2(e). It can be seen from a comparison between Figs.
2(a) and 2(c) that when the tip distance $d_y$ decreases from
16.07 \AA\ to 15.27 \AA, the conduction band of the deformed
(10,0) tube shifts down to the Fermi level and tends to split off,
but the valence band is still similar to that of the perfect tube,
accompanied by a very small energy split, which leads the energy
gap to be reduced. Of special interest is that the non-degenerate
state labeled by (s) in Fig. 2 moves down under radial
deformation, which is caused by the enhanced $\sigma^*-\pi^*$
hybridization, since the tube curvature is increased in the curved
tube regions [20,21,22]. When the $d_y$ decreases to 14.47 \AA,
the gap is closed due to the lowest split conduction band touching
the highest valence band, indicating a phase transition from
semiconductor to metal (see Fig. 2(d)). Further squashing the DWNT
($d_{y}=$ 13.67 \AA, Figs. 1(c) and 1(d)) will make the lowest
split conduction band shift down deeper, leading to intercross of
the original conduction and valence bands (see Fig. 2(e)), and
meanwhile, all the bands near the Fermi level to be
non-degenerate. In order to check the validity of the TB
calculations, we have also carried out the first-principles
calculation [29] on the deformed inner (10,0) tube with $d_{y}=$
13.67 \AA. The obtained results (shown in Fig. 2(f)) indicate
that, our TB calculations can correctly reproduce the
first-principles results.

Next, we investigate the transport properties of the MSM DWNT IMJ
by the above said method. Since the outer tube actually acts as an
insulator coat of the inner metal wire, we do not take into
account it in our transport calculation.

As is well known, the ballistic conductance at an energy level is
proportional to the number of conduction channels at the same
energy [30], from which the change of band structure may be
clearly found by the conductance measurement. We have calculated
the conductance and the total density of states (DOS) of the MSM
DWNT IMJ with the central semiconductor length of $L=$ 14 unit
cell, and obtained results are given in Figs. 3(a) and 3(b). The
calculated conductances of the perfect semiconducting (10,0) tube
and the squashed metal (10,0) tubes ($d_{y}=$ 13.67 \AA) are also
shown in Fig. 3(a).

It is seen from Fig. 3(a) that radical squash induces the
significant changes of the conductance near the Fermi level. And
the conductances of the MSM DWNT IMJ below the Fermi level and
above 1.37 eV are similar to those of deformed metal (10,0) tube.
The biggest changes of the conductance appear in the energy window
from 0 to 1.37 eV. Firstly, at 0.52 eV there is a conductance
jump, which arises from the conductance jump of the squashed metal
(10,0) tube at the 0.52 eV. Secondly, in the region of 0 $\sim$
0.69 eV (the band gap of the ideal (10,0) tube), there exists
nonzero conductance, which comes from quantum tunnelling. Third,
the resonance electron transmissions occur in the energy area of
0.69 $\sim$ 1.37 eV, giving three conductance oscillation peaks at
0.69, 0.88 and 1.07 eV, respectively. Using a 1D square potential
barrier with finite height, which is used to model the central
semiconducting segment of the DWNT IMJ, we can simply get three
oscillation peaks in their positions to be the same as those of the above three
conductance peaks.

These interesting phenomena related to the conductance can be
explained by the DOS shown in Fig. 3(b). From Fig. 3(b), we can
see that there are four DOS peaks in the energy area of 0 $\sim$
1.37 eV, coinciding well with the four conductance peaks in the
same energy window of Fig. 3(a). The LDOS along the tube axis at
these four energy levels of 0.52, 0.69, 0.88 and 1.07 eV are shown
in Fig. 4(a), it is clearly seen that all of the four states are
extended along the tube axis. Also, the LDOS on the up surface of
the MSM DWNT IMJ, shown in the left panel of Fig. 4(a), is much
lower than that on the side surface, shown in the right panel,
indicating that the conductances at these energy levels are
contributed mostly by the atoms on the side surface [31]. In Fig.
4(b), we show the LDOS at four energy levels of -0.05, -0.09,
-0.27 and -0.46 eV in valence band, respectively. The large
contribution to conductance, coming from the atoms on the side
surface of the MSM DWNT IMJ, can also be seen from Fig. 4(b). But
we found that at energy levels of -0.05 and -0.09 eV exist the
quasibound states with their largest LDOS at positions of $\pm
34.43$ \AA\ (represented by the broken vertical lines) lying in
the transition region between the deformed inner metal tube and
the semiconductor tube. However, LDOSs at the energy levels of
-0.27 and -0.46 eV show the characteristics of extended states.

In order to know effect of the length of central semiconducting
part in the MSM DWNT IMJ, we have also calculated conductance of
the IMJ with $L=$ 8 unit cell and shown the obtained result in
Fig. 3(a). For the transmission electrons, the middle
semiconductor act as an energy barrier. And it is well known that
the tunnelling probability decays exponentially with the
increasing width of the energy barrier, which is well consistent
with what shown in Fig. 3(a). Comparing with the conductance of
the IMJ with $L=$ 14 unit cell, we find the lower edge of the
energy gap shifts to a higher energy, and a bigger conductance
jump to $G_0 $ (where $G_0 = 2e^2 / h$ is the conductance quantum)
appears at 0.52 eV due to the stronger quantum tunnelling. And in
the region of 0.69 $\sim$ 1.37 eV, the resonance electron
transmission peaks move to the higher energies.

Based on the MSM DWNT IMJ, we can easily make a pure carbon
nanoscale transistor by applying a gate voltage $V_{g} $ to the
DWNT junction with a inner semiconductor (10,0) tube. When $V_{g}
=$ 0, the conductance is zero and the transistor is OFF. With
change of $V_{g} $, the transistor can jump to an ON state.

\section {Conclusion}

In summary, based on Slater-Koster tight-binding and the
first-principle calculations, we found that the different
radically squashed DWMT IMJs have very different $\sigma-\pi$ and
$\sigma^*-\pi^*$ hybridizations, inducing big changes of their
electronic structures. It has been shown that the big enough
radial squash can induce an insulator-metal transition for the
inner semiconducting (10,0) tube, but the outer (19,0) tube
remains to be semiconducting. The LDOSs of the MSM DWNT IMJ at
different energies indicate the emergence of two quasibound states
at -0.05 and -0.09 eV near the Fermi level, and large
contributions to the conductance, coming from the atoms on the
side surfaces of the IMJ. In addition, There exist several
resonance conductance peaks at some energies in the conduction
band of the MSM DWNT IMJ, which can be explained by electron
resonance transmission effect. Also, the conductance variation due
to length change of the central semiconducting part in the MSM
DWNT IMJ has been found. Finally, a promising pure carbon
nanoscale transistor is proposed based on the DWNT IMJ.

\begin{acknowledgments}

We acknowledge discussions with Gang Wu. This work is supported by
the Natural Science Foundation of China under Grants No. 90103038
and No. A040108.
\end{acknowledgments}

\newpage
\begin{center}
\textbf{Figure Captions}
\end{center}

\begin{figure}[htbp]
\includegraphics[width=0.8\columnwidth]{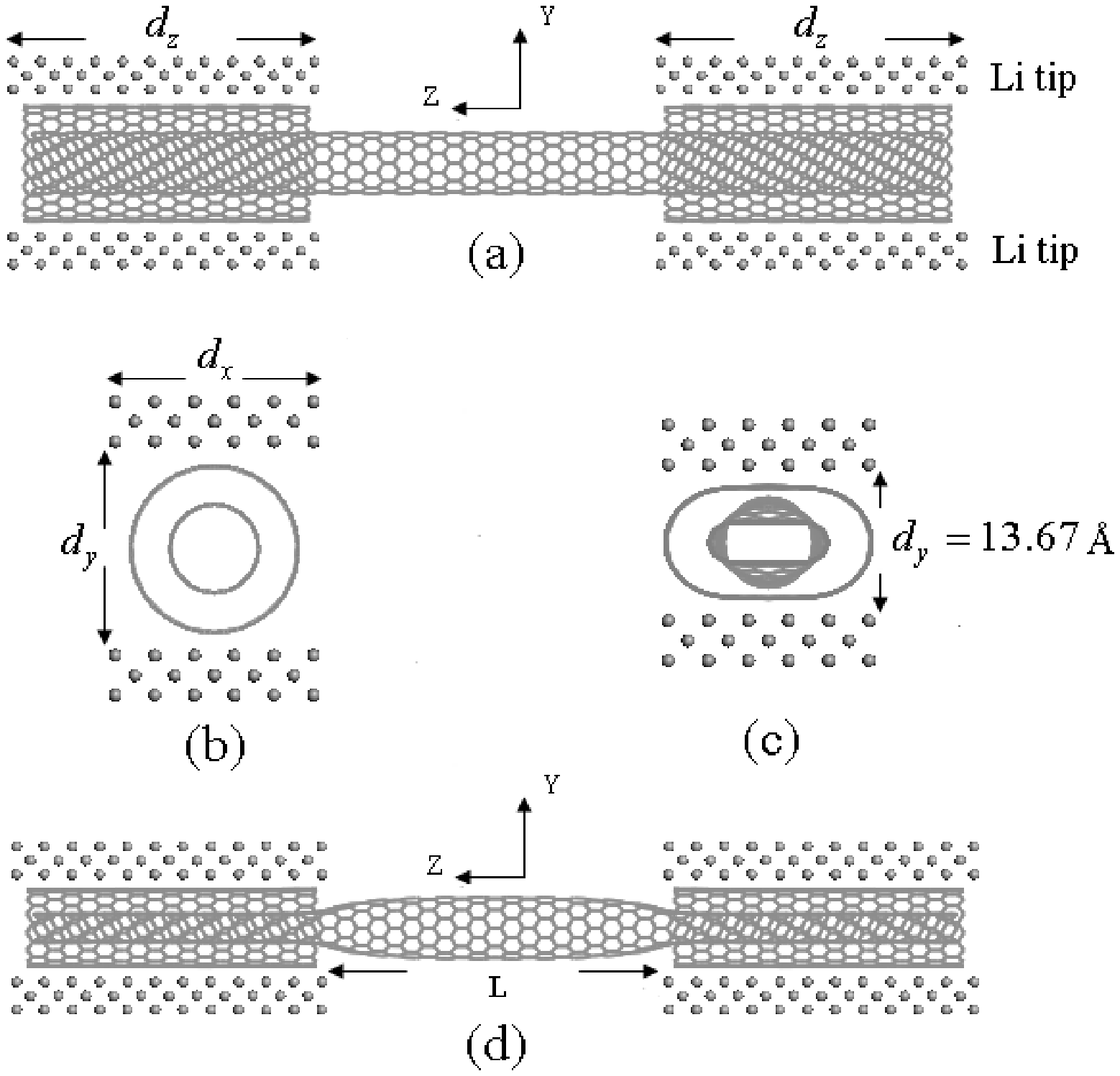}
\label{fig1} \caption{Schematic view of a finite (10,0)@(19,0)DWNT
with a central segment of uncovered inner tube along (a) $x$-axis
direction, and (b) $z$-axis direction, respectively. (c) and (d)
Relaxed atomic structures of the deformed (10,0)@(19, 0) DWNT by the
Li tip with separation distance $d_{y}$ of 13.67 \AA. $L$ is the
length of the uncovered semiconducting inner (10,0) tube in the DWNT
IMJ.}
\end{figure}

\begin{figure}[htbp]
\includegraphics[width=0.8\columnwidth]{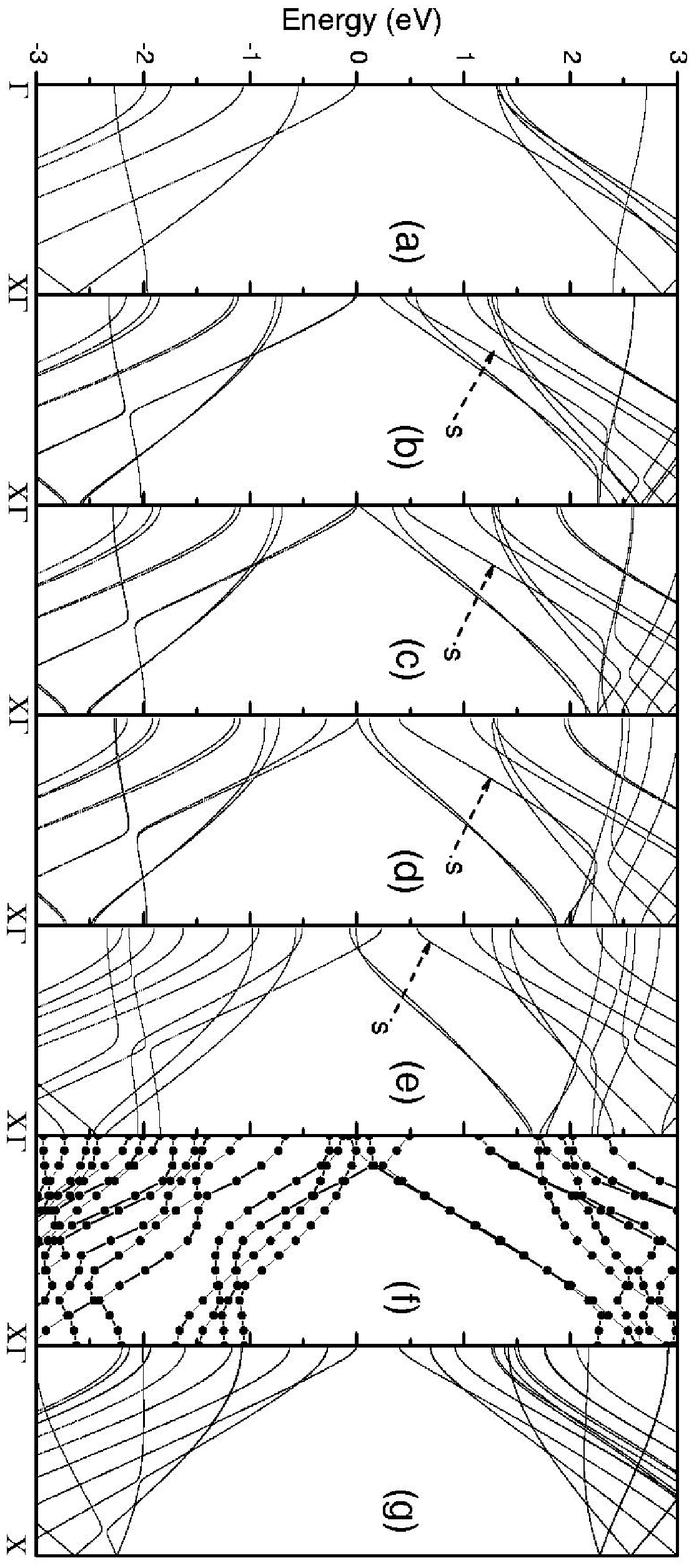}
\label{fig2} \caption{Band structures of (a) perfect (10,0) tube and
(b), (c), (d), (e) deformed (10,0) tubes with different $d_{y}$
values of 16.07 \AA, 15.27 \AA, 14.47 \AA, and 13.67 \AA,
respectively. (f) The band structure obtained by the
first-principles calculation for the radially deformed (10,0) tube
with $d_{y}$ of 13.67 \AA. (g) The band structure of the deformed
(19,0) tube with $d_{y}$ values of 13.67 \AA. The Fermi level is set
at zero.}
\end{figure}

\begin{figure}[htbp]
\includegraphics[width=0.8\columnwidth]{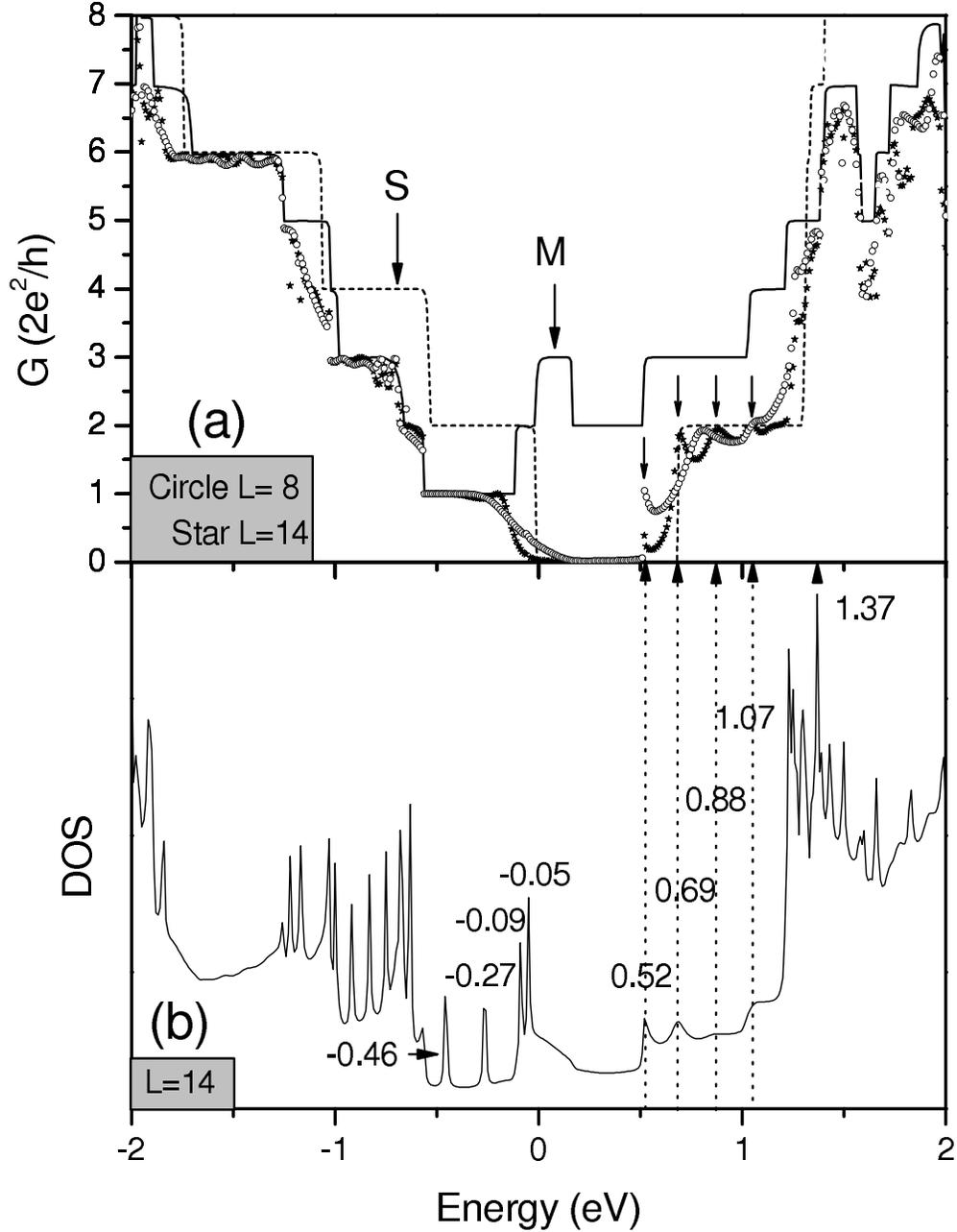}
\label{fig3} \caption{(a) Conductance of the perfect semiconductor
(10,0) tube, the radially deformed metal (10,0) tube with  $d_{y}$
of 13.67 \AA\ (represented by the dash and solid line,
respectively), and that of the MSM DWNT IMJ with two different
semiconductor lengths of $L = 8$ unit cell (the Circle), and $L =
14$ unit cell (the star). (b) The total DOS of the IMJ with $L=$ 14
unit cell. The Fermi level is set at zero.}
\end{figure}

\begin{figure}[htbp]
\includegraphics[width=0.8\columnwidth]{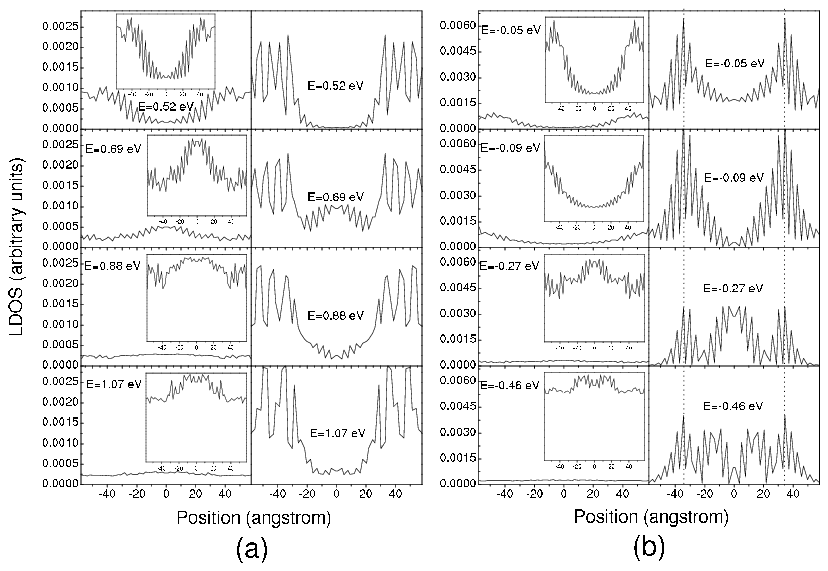}
\label{fig4} \caption{The LDOSs of the MSM DWNT IMJ at different
energy levels with its middle point set at zero. (a) At four energy
levels of 0.52, 0.69, 0.88 and 1.07 eV in the conduction band. (b)
At four energy levels of -0.05, -0.09, -0.27 and -0.46 eV in the
valence band. In (a) and (b), the left and right panels correspond,
respectively, to a scanning parallel to the tube axis along the
atom-line on the up surface ($y > 0.0$ \AA, $\left| x \right| < 0.6
$ \AA) , and on the side surface ($x > 0.0$ \AA, $\left| y \right| <
0.6 $ \AA). The insets in the left panels show the corresponding
enlarged LDOSs.}
\end{figure}


\begin{thebibliography}{99}


\bibitem {r1} S. Iijima, Nature (London) {\bf 354}, 56 (1991).

\bibitem {r2} T.W. Ebbesen, P.M. Ajayan, Nature (London) {\bf 358},
220 (1992).

\bibitem {r3} J. Kong, H.T. Soh, A.M. Cassell, C.F. Quate, H. Dai, Nature (London) {\bf 395},
878 (1998).

\bibitem {r4} R. Saito, G. Dresselhaus, and M.S. Dresselhaus,
Physical Properties of Carbon Nanotubes (Imperial College Press,
London, 1998).

\bibitem {r5} H.J. Dai, A.G. Rinzler, P. Nikolaev, A. Thess, D.T. Colbert, R.E.
Smalley, Chem. Phys. Lett. {\bf 260}, 471 (1996).

\bibitem {r6} R. Saito, R. Matsuo, T. Kimura, G. Dresselhaus, M.S. Dresselhaus, Chem.
Phys. Lett. {\bf 348}, 187 (2001).

\bibitem {r7} K. Tanaka, H. Aoki, H. Ago, T. Yamabe, K. Okahara, Carbon {\bf 35},
121 (1997).

\bibitem {r8} Y.K. Kwon, D. Tomanek, Phys. Rev. B {\bf 58},
R16001 (1998).

\bibitem {r9} M. Buongiorno Nardelli, C. Brabec, A. Maiti, C. Roland, J. Bernholc,
Phys. Rev. Lett. {\bf 80}, 313 (1998).

\bibitem {r10} T.W. Tombler, C. Zhou, L. Alexseyev, J. Kong, H. Dai,
L. Liu, C.S. Jayanthi, M. Tang, and S. Wu, Nature (London) {\bf
405}, 769 (2000).

\bibitem {r11} E.D. Minot, Y. Yaish, V. Sazonova, J.Y. Park, M. Brink,
and P.L. McEuen, Phys. Rev. Lett. {\bf 90}, 156401 (2003).

\bibitem {r12} J. Cao, Q.Wang, and H. Dai, Phys. Rev. Lett. {\bf 90}, 157601
(2003).

\bibitem {r13} A. Maiti, A. Svizhenko, and M.P. Anantram, Phys. Rev.
Lett. {\bf 88}, 126805 (2002).

\bibitem {r14} M.B. Nardelli and J. Bernhole, Phys. Rev. B {\bf 60}, R16338
(1999).

\bibitem {r15} J.Q. Lu, J. Wu, W. Duan, F. Liu, B.F. Zhu, and B.L.
Gu, Phys. Rev. Lett. {\bf 90}, 156601 (2003).

\bibitem {r16} J.Q. Lu, J. Wu, W. Duan, and B.L. Gu,
arxiv:cond-mat/0312655.

\bibitem {r17} Susumu Okada and Atsushi Oshiyama, Phys. Rev. Lett.
{\bf 91}, 216801 (2003).

\bibitem {r18} Pingan Hu, Kai Xiao, Yunqi Liu, Gui Yu, Xianbao
Wang, Lei Fu, Guanglei Cui, and Daoben Zhu, Appl. Phys. Lett. 84,
4932 (2004).

\bibitem {r19} A.K. Rappe, C.J. Casewit, K.S. Colwell, W.A.
Goddard, W.M. Skiff, J. Am. Chem. Soc. {\bf 114}, 10024 (1992).

\bibitem {r20} N. Yao, V. Lordi, J. Appl. Phys. {\bf 84}, 1939 (1998).

\bibitem {r21} J.C. Charlier, Ph. Lambin, and T. W. Ebbesen,
Phys. Rev. B {\bf 54}, 8377 (1996).

\bibitem {r22} H.S. Sim, C.J. Park, and K. J. Chang,
Phys. Rev. B {\bf 63}, 073402 (2001).

\bibitem {r23} X. Blase, L.X. Benedict, E.L. Shirley, and S. G. Louie,
Phys. Rev. Lett. {\bf 72}, 1878 (1994).

\bibitem {r24} J.C. Slater and G.F. Koster, Phys. Rev. {\bf 94}, 1498 (1954).

\bibitem {r25} X.P. Yang, J.W. Chen, H. Jiang and J. Dong, Phys. Rev. B {\bf
69}, 193401 (2004).

\bibitem {r26} D. Tomanek and S.G. Louie, Phys. Rev. B {\bf 37}, 8327 (1988).

\bibitem {r27} M.B. Nardelli and J. Bernholc, Phys. Rev. B {\bf 60}, 7828 (1999).

\bibitem {r28} C. Caroli, R. Combescot, P. Nozieres, and D. Saint-James,
J. Phys. C: Solid St. Phys. {\bf 4}, 916 (1971).

\bibitem {r29} M.D. Segall \textit{et al.}, J. Phys.: Cond. Matt. {\bf 14}, 2717
(2002)

\bibitem {r30} D.S. Fisher and P.A. Lee, Phys. Rev. B {\bf 23}, 6851 (1981).

\bibitem {r31} J.W. Chen, X.P. Yang, L.F. Yang, H.T. Yang and J.
Dong, Phys. Lett. A {\bf 325}, 149 (2004).



\end{thebibliography}
\end{document}